\documentclass[letter,scriptaddress,twocolumn, prl]{revtex4}

	\usepackage{amsmath}
	\usepackage{makeidx}
	\usepackage{amsfonts}
	\usepackage[ansinew]{inputenc}
	\usepackage[usenames,dvipsnames]{pstricks}
	\usepackage{subfigure}
	\usepackage{epsfig}
	\usepackage{epstopdf}
	\usepackage{braket}
	\usepackage{pst-grad} 
	\usepackage{pst-plot} 
	\usepackage[colorlinks,hyperindex]{hyperref}
	\hypersetup
	{
		colorlinks,%
		citecolor=black,%
		linkcolor=black,%
		urlcolor=black,%
	}
\makeindex
\newcommand\numberthis{\addtocounter{equation}{1}\tag{\theequation}}

\begin{document}

\title{Multislip Friction with a Single Ion}
\author{Ian Counts,$^{1}$ Dorian Gangloff,$^{1,2}$ Alexei Bylinskii,$^{1,3}$ Joonseok Hur,$^{1}$ Rajibul Islam,$^{1,4}$ and Vladan Vuleti\'{c}$^{1,}$}
\email{vuletic@mit.edu}
\affiliation{
$^{1}$Department of Physics and Research Laboratory of Electronics, Massachusetts Institute of Technology, Cambridge, Massachusetts 02139, USA \\
$^{2}$Cavendish Laboratory, JJ Thompson Ave, Cambridge CB3 0HE, UK \\
$^{3}$Department of Chemistry and Chemical Biology and Department of Physics, Harvard University, Cambridge, Massachusetts 02138, USA \\
$^{4}$Institute for Quantum Computing and Department of Physics and Astronomy, University of Waterloo, Waterloo, Ontario N2L 3G1, Canada 
}

\begin{abstract}
A trapped ion transported along a periodic potential is studied as a paradigmatic nanocontact frictional interface.  The combination of the periodic corrugation potential and a harmonic trapping potential creates a one-dimensional energy landscape with multiple local minima, corresponding to multistable stick-slip friction.  We measure the probabilities of slipping to the various minima for various corrugations and transport velocities.  The observed probabilities show that the multislip regime can be reached dynamically at smaller corrugations than would be possible statically, and can be described by an equilibrium Boltzmann model.  While a clear microscopic signature of multislip behavior is observed for the ion motion, the frictional force and dissipation are only weakly affected by the transition to multistable potentials.
\end{abstract}
\maketitle

\begin{figure}[!htbp]
	\begin{center}
		\includegraphics[width=0.37\textwidth]{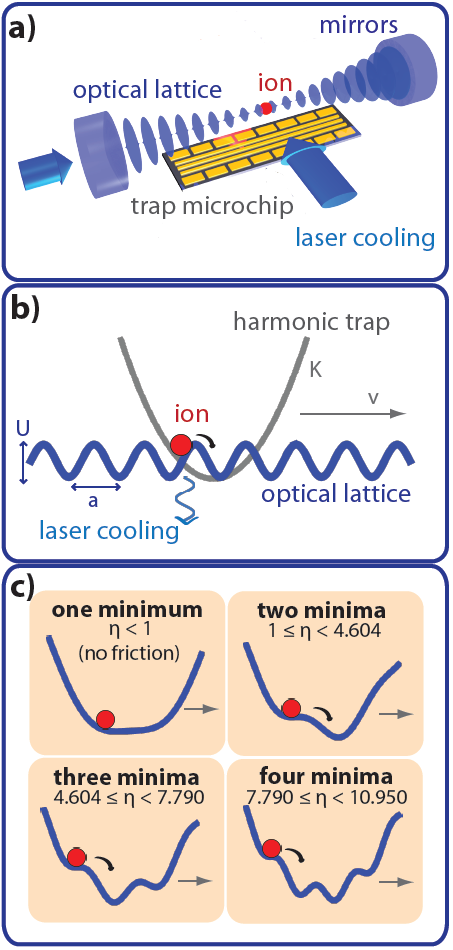} 
		\caption{\textbf{a)} Schematic of the experimental set-up.  A $^{174}$Yb$^+$ ion is trapped by a linear Paul trap, located 135 $\mu$m above the surface of a lithographic microchip.  The chip generates radial trapping with an RF field and axial trapping with a DC harmonic potential.  An axial optical-lattice standing wave is produced by 370 nm light, blue-detuned from the $^2S_{1/2} \rightarrow$ $^2P_{1/2}$ atomic transition by 12.6 GHz.  Ion fluorescence is collected during laser cooling. \textbf{b)} Stick-slip friction in a periodic optical potential.  As the harmonic trap is translated at speed $v$, it drags the ion along the sinusoidal optical-lattice potential, causing the ion to slip.  The excess energy acquired during a slip is dissipated by continuous laser cooling.    \textbf{c)} Combined harmonic-sinusoidal potential for different values of the corrugation parameter $\eta$. The potential energy landscape is drawn just before the slipping point, where the left-most minimum vanishes and a potential with initially $n$ local minima has $n-1$ available minima to which the ion can slip.  Note that for $n=1$, there is no energy barrier and, consequently, no stick-slip friction.}
	\end{center}
\end{figure}

Stick-slip friction is a ubiquitous non-equilibrium dynamical process that occurs at the interface between surfaces across a wide range of length scales \cite{scholz,urbakh,bormuth,mo,urbakh2,vanossi}.  The term \emph{stick-slip} describes the system's response to an applied shear force: the surfaces slip out of a local minimum in the interface energy landscape, and stick into a new lower-energy minimum, releasing heat in the process.

Recent advances in atomic force microscopy (AFM) have extended the study of stick-slip friction to the atomic scale, where atom-by-atom slips occur at the interface between a probe tip and a periodic substrate \cite{binnig,carpick,carpick2,krylov,meyer,gnecco,riedo,liu,li,zhao,jansen,barel}.  For a single-atom probe, the number of local minima in the probe-substrate interaction potential is determined by the ratio of the periodic substrate potential to the spring constant with which the probe is bound to its support object.  As the load on the probe is increased, or equivalently, the periodic substrate potential is deepened, the system transitions from a bistable regime (where the probe deterministically single-slips from the first minimum to the second) to a multistable regime (where a probe can stochastically multislip to one of several local minima).  This has been demonstrated in AFM simulations \cite{nakamura,dong,tshiprut,evstigneev} and experiments \cite{mate,medyanik,roth} where single-slip and multislip events have been clearly differentiated.  However, in the absence of control over dissipation rates and the microscopic energy landscape, it is difficult to tie the observations to ab initio friction models.

Following theoretical proposals \cite{tosatti,tosatti2,shepelyansky,haffner}, we have recently demonstrated a trapped-ion friction emulator with extensive control over all microscopic interface parameters \cite{science,velocity,aubry}.  In analogy to AFM, the emulator features a small probe (one or several trapped ions) transported over a periodic substrate potential created by an optical standing wave (Figs.~1a,b) \cite{enderlein,linnet,karpa}.  To date, we have used the emulator to study the velocity-dependence of nanofriction \cite{velocity}, as well as the interplay between superlubricity \cite{science,shinjo,hirano,dienwiebel,socoliuc} and the Aubry transition \cite{aubry}.  These studies and a recent study of zig-zag ion chains \cite{mehlstaubler} have focused on the single-slip regime.

In this Letter, we study multislip friction in deep substrate potentials.  We observe the ion fluorescence associated with slip events, from which we directly extract the temperature- and velocity-dependent probabilities for the ion to localize in one of the available local minima.  We find that at finite rethermalization times following a slip, the multislip regime can be reached dynamically at smaller corrugations than would be possible statically.  We also find that the probabilities agree well with a simple Boltzmann model, despite the dynamical nature of the process.  Remarkably, the average frictional energy dissipation $U_{diss}$ and the maximal static friction force $F_{static}$ are mostly unaffected by the transition from the single-slip to the multislip regime, increasing approximately linearly with the depth of the substrate potential.

The potential energy landscape experienced by the ion is produced by the combination of an electrostatic harmonic potential provided by a linear Paul trap \cite{cetina} and a sinusoidal optical lattice \cite{karpa,science,velocity,aubry}.  The potential energy of the ion at position $x$ is given by the Prandtl-Tomlinson model \cite{P,T}:
\begin{equation}
\frac{V(x)}{Ka^2}=\frac{1}{2}\left(\frac{x-x_0}{a}\right)^2+\frac{\eta}{4\pi^2}\left(1-\cos\left(\frac{2\pi x}{a}\right)\right).
\end{equation}
The first term is attributed to the harmonic trap at position $x_0$, corresponding to a spring with constant $K=m \omega_0^2$ ($m$ is the mass of the $^{174}$Yb$^{+}$ ion and $\omega_0/(2\pi) \approx 360$ kHz is the axial vibrational frequency of the harmonic trap).  The second term is due to the AC Stark shift of the lattice with period $a = 185$ nm \cite{science}.  The number of local minima in $V(x)$ is determined by the corrugation parameter $\eta= (\frac{\omega_L}{\omega_0})^2$, with $\omega_L$ the vibrational frequency at the lattice minima.  By adjusting the optical-lattice amplitude to a maximum of $U/h = 40$ MHz, we change $\omega_L = \sqrt{\frac{2 \pi^2 U}{m a^2}}$ up to $2\pi\times 1.1$ MHz, and thus tune $\eta$ in the range $0 \le \eta \le 10$.  Values of interest include $\eta=1, 4.604,$ and $7.790$, which mark the transition to potentials with $n=2$, $n=3$, and $n=4$ local minima, respectively (Fig.~1c).

The ion is transported by adjusting the potentials of the trap electrodes so as to translate the harmonic-trap position $x_0$.  Thus, we drive the ion over the optical lattice at constant average velocity $v = \frac{dx_0}{dt}$, forcing the ion to slip over lattice maxima (Fig.~1c).  During the transport, the ion is continuously laser cooled via Raman sideband cooling to a typical temperature of $50$ $\mu$K ($k_BT/U$ in the range $0.5$ to $0.03$ for $\eta$ in the range $0.5$ to $10$) \cite{karpa}, and observed via the fluorescence emitted during the cooling process.  For a stationary ion, the fluorescence peaks when its stable minimum becomes an inflection point, the moment when the ion is closest to the maximum of the optical-lattice potential \cite{karpa}.  After the ion slips over a lattice maximum, its fluorescence falls exponentially while it cools and localizes into a new local minimum.

\begin{figure}[!htbp]
	\begin{center}
		\includegraphics[width=0.45\textwidth]{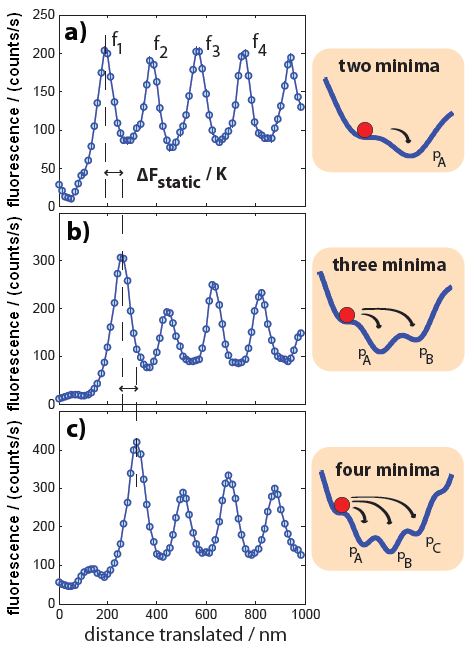} 
		\caption{Measured ion fluorescence as a function of trap translation, indicating single-slip vs. multislip behavior.  After the ion is prepared in the global energy minimum, the harmonic trap is translated at constant velocity across the optical standing wave.  This forces the ion to slip over local potential maxima, resulting in fluorescence peaks. The fluorescence traces are averaged over an exposure time of 300 seconds, corresponding to roughly $1.5 \times 10^{5}$ realizations of the experiment.  Error bars are statistical and indicate one standard deviation.  \textbf{a)} Fluorescence trace indicating single-slip behavior for a potential with $n=2$ minima ($\eta = 3.2$).  Fluorescence peaks are of equal height, indicating that the ion always slips to the adjacent minimum (lattice period $a = 185$ nm).  There is a small variation of peak heights due to the finite recooling time (see text).  \textbf{b,c)} Fluorescence traces indicating multi-slip behavior in a potential with $n=3$ and $n=4$ minima, respectively ($\eta = 6.4$ and $\eta = 9.6$). Fluorescence peaks vary in height, indicating that an ion can slip into one of multiple local minima, and therefore may not always slip when a given minimum disappears.  Note the hysteretic delay of the peaks in these time traces compared to Fig.~2a.  This is due to the greater static friction force $F_{static}$ exerted on the ion by the optical lattice at higher $\eta$.  (See Ref. \cite{science} and Fig.~4a.)}  
	\end{center}
\end{figure}

Initializing the ion in the global potential minimum, and then transporting through consecutive slip events, we observe a series of fluoresence peaks; the relative heights of these peaks differentiate single-slip from multislip behavior.  For an ion undergoing single-slips, a series of equally-spaced fluorescence peaks of equal height is observed, as the ion always localizes in the adjacent minimum after every slip event (Fig.~2a).  The transition to the multislip regime manifests itself as fluorescence peaks of different heights, associated with random localizations in more distant minima (Fig.~2b,c).  

To see why the two slip modes result in different peak height distributions, we note that the fluorescence traces are averaged over multiple repetitions of the initialization-transport experiment.  The more likely an ion is to slip at a particular time, the higher the associated averaged fluorescence peak.  The ion is initialized in the global potential minimum; when this minimum vanishes due to trap translation, the ion will \emph{always} slip and fluoresce.  Thus, the first peak $f_1$ is the largest.  After this initial slip, if the ion localizes in the adjacent minimum (single-slip), then further trap translation by one lattice period $a$ will cause the ion to slip and fluoresce again, and we will observe $f_2 = f_1$.  If it localizes instead in the next-adjacent minimum (multislip), then a fluorescence peak will not appear until translation by $2a$.  A finite probability of next-adjacent localization will result in a reduced peak $f_2 < f_1$ and a higher peak $f_3 > f_2$ (see Fig.~2b,c).  

The relationship between the localization probabilities $\{p_A,p_B,p_C\}$ (where $p_A$ denotes localization in the adjacent minimum, $p_B$ the next-adjacent minimum, etc.) and the peak height distribution $\{f_i\}$ is given by \cite{supplemental}:
\begin{align*}
p_A &= f_2 / f_1 \\ \numberthis \label{eqn}
p_B &= - (f_2 / f_1)^2 + f_3 / f_1 \\
p_C &= (f_2 / f_1)^3 - 2 f_2 f_3 / f_1^2 + f_4/f_1.
\end{align*}
Note that the probability distribution is extracted directly from the observed peak heights without making assumptions about the localization process.

\begin{figure}[!htbp]
	\begin{center}
		\includegraphics[width=0.46\textwidth]{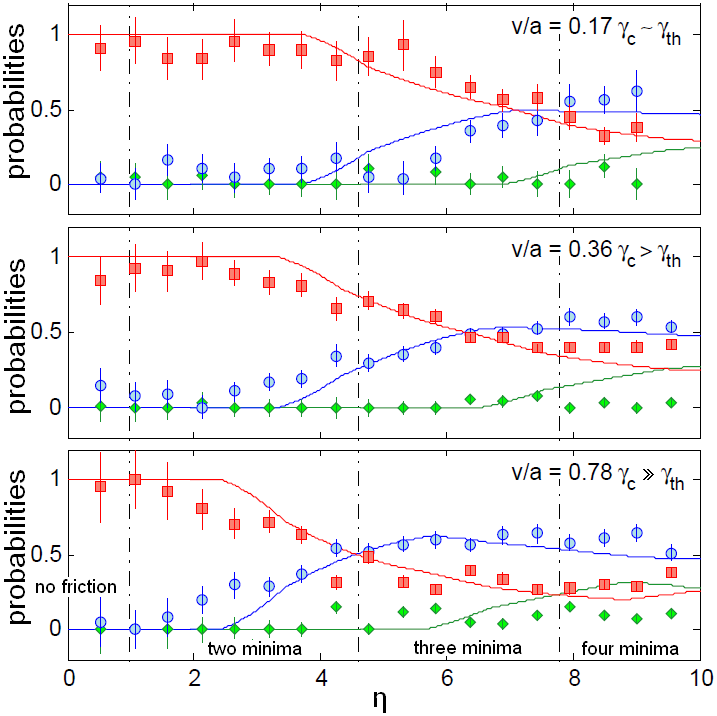} 
		\caption{Multislip probabilities vs. corrugation parameter $\eta$ for different transport velocities $v$.  Data points are the extracted slipping probabilities $p_A$ (slip to next minimum, red squares), $p_B$ (slip to second-next minimum, blue circles), $p_C$ (slip to third minimum, green diamonds).  Velocities are reported as functions of the system's recooling rate $\gamma_c$ and thermal hopping rate $\gamma_{th}$, the rate at which the ion hops over a barrier due to its finite temperature (see main text).  Error bars are statistical and indicate one standard deviation.  The multislip regime is distinguished by non-negligible values of $p_B$.  Curves are calculated from a theoretical model.  Vertical dash-dotted lines separate regimes with different numbers of minima (as in Fig.~1c).}
	\end{center}
\end{figure}

By measuring fluorescence patterns like the ones shown in Fig.~2 for different corrugation parameters $\eta$ and transport velocities $v$, we extract localization probabilities over a range of experimental conditions (Fig.~3).  For all three velocities shown, multislip behavior (second-next neighbor slip probability $p_B > 0$) is observed in the multistable regime ($\eta > 4.604$).  This is a consequence of the underdamping of the system ($\gamma_c \ll \omega_0$), which guarantees that an ion, following a slip, can sample the full potential landscape before it recools at rate $\gamma_c = 10^4$ s$^{-1}$ and localizes in a minimum.  More surprising is the appearance of multislip events before the corrugation is deep enough to create multiple static minima (in the region of bistable potential $\eta < 4.604$, where a third minimum should not yet exist).  This is the case for the two fastest transport speeds but not for the slowest transport.  This can be readily explained by the change in the energy landscape during the recooling time $\gamma_c^{-1}$: by the time the ion is sufficiently cooled to localize, another potential minimum may have opened up if $v/a \gtrsim \gamma_c$.  Thus, we find that the multislip regime can be reached dynamically at smaller corrugations than would be possible statically.

The experimental data shown in Fig.~3 is overlayed with a theoretical model that takes into account the system's competing rates (transport rate $v/a$ and recooling rate $\gamma_c$).  Our model's central assumption is that an ion is more likely to localize in a lower-energy minimum and that this effect can be described by a quasi-equilibrium Boltzmann probability $p_i \propto \exp(-\frac{V_i(\tau)}{k_B T(\tau)})$.  Here, $V_i(\tau)$ is the potential energy of the $i$th minimum at time $\tau$ when the ion localizes, and $T(\tau)$ is the temperature of the ion at that time.  To model its temperature, we note that an ion has some potential energy $V_0$ at the slipping point.  This is converted into kinetic energy and dissipated exponentially by laser cooling: $k_B T(\tau) = V_0 \exp(-\gamma_c \tau)$.  Our model's free parameter is the localization time, found to be $\tau = (65 \pm 5$) $\mu$s by fitting the model to the data.

We note that the dataset with the slowest transport speed is fitted with lower confidence by the model above $\eta > 4.604$.  This discrepancy is the signature of another dynamical rate of the system, the thermal hopping rate $\gamma_{th} = 10^3$ s$^{-1}$, observed previously in Refs. \cite{velocity,jinesh,sang}.  Thermal hopping across a barrier due to the ion's finite temperature dominates at the slowest transport speed, where $v/a \sim \gamma_{th}$ (the thermolubric regime \cite{velocity}).  Its effect on the fluorescence signal is to smooth the peak height distribution, which causes us to overestimate the value of $p_A$ and underestimate $p_B$ \cite{supplemental}.  Evidently, the relationship between fluorescence and probability (Eq.~2) is strictly valid only for faster transport speeds, where thermal hopping is negligible ($v/a \gg \gamma_{th}$).

Thermal hopping also affects the observed slip probability to the third minimum $p_C$, which for large $\eta$ and large speed is distinctly smaller than predicted.  The third minimum, most distant from the slipping point, has the smallest potential barrier and the longest dwell time before the minimum disappears, making the ion most susceptible to thermal hopping out of that minimum, even at fast transport speeds.  Because a thermal hop at a random time does not result in an (averaged) fluorescence peak, we undervalue the localization probability $p_C$.

\begin{figure}[!htbp]
	\begin{center}
		\includegraphics[width=0.46\textwidth]{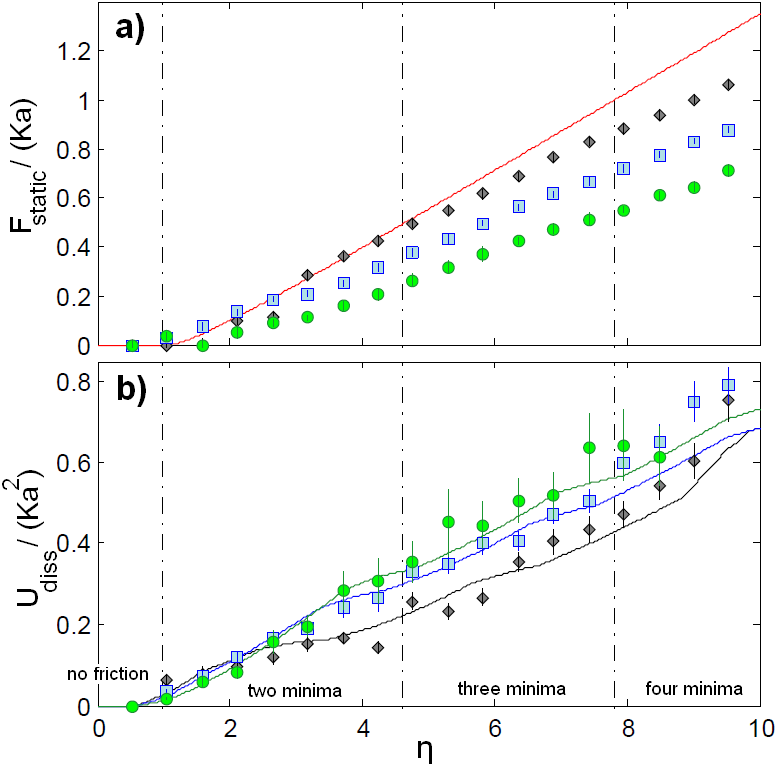} 
		\caption{\textbf{a)} Measured static friction force vs $\eta$ for three different drive velocities ($v/a = 0.17$ $\gamma_c$ for green circles, $v/a = 0.36$ $\gamma_c$ for blue squares, $v/a = 0.78$ $\gamma_c$ for black diamonds).  The velocity-dependence of the slope is due to thermal hopping, which reduces friction for slow transport \cite{velocity}.  The red theory curve is a realization of the Prandtl-Tomlinson model with no free parameters, without thermal hopping.  \textbf{b)} Energy dissipated per slip vs $\eta$.  Data points are calculated from experimentally determined values for $p_i$, while the curves use the model values for $p_i$ from Fig.~3.  Both use the calculated energy landscape.  Vertical dash-dotted lines separate regimes with different numbers of minima.} 
	\end{center}
\end{figure}

Figs.~2 and 3 show clear delineations of single-slip and multislip behavior as a function of $\eta$.  Interestingly, purely frictional quantities, like the maximal static friction force $F_{static}$ and energy dissipated per slip $U_{diss}$, do not reveal clear signatures of the transition.  Fig.~4a shows that $F_{static}$ exerted by the lattice on the ion increases monotonically with the corrugation parameter $\eta$, without any discernable changes near the critical values for multistability ($F_{static}$ is determined from the observed hysteretic shift of the fluorescence peaks as a function of trap translation, see Fig.~2 and Ref. \cite{science}).  Fig.~4b shows the calculated average energy $U_{diss}$ dissipated per slip by laser cooling, determined from the measured slip probabilites $p_i$ and the calculated energy landscape $V(t)$ \cite{supplemental}.  Like $F_{static}$, the dissipated energy $U_{diss}$ increases monotonically with the corrugation parameter $\eta$, and shows little dependence on the number $n$ of potential minima in the energy landscape for $n \geq 2$.  This can be attributed to the cancelation of two competing effects: if an ion slips to the second-next minimum rather than to the next minimum, it will release more heat, as the second-next minimum is lower in energy by the time the ion localizes.  On the other hand, an ion slipping to a more distant minimum will wait longer before it slips again.  Thus, to leading order, the dissipation is independent of the slipping mode.  To higher order, the slope $dU_{diss}/d\eta$ is slightly reduced for a potential with more minima, as the reduction in slip frequency overpowers the smaller increase in dissipated energy per slip.  This trend is visible in Fig.~4b, where the slope is reduced at the transition between single-slip and multislip behavior.

In this work, we have measured the slip probabilities for stick-slip friction in a multistable energy landscape, and have shown that the dynamic stick-slip process can be described by a quasi-equilibrium model.  We suggest that the model's predictive power could be used in nanopositioning applications: by tuning system parameters, a probe could be engineered to multislip to a specific potential minimum. In the future, these studies could be extended to the quantum regime in order to study quantum annealing in a multistable energy landscape.

This work was supported in part by the NSF, the NSF CUA, and the MURI program through ONR.

\end{document}


\title{Multislip Friction with a Single Ion - Supplemental Material}

\author{Ian Counts,$^{1}$ Dorian Gangloff,$^{1,2}$ Alexei Bylinskii,$^{1,3}$ Joonseok Hur,$^{1}$ Rajibul Islam,$^{1,4}$ and Vladan Vuleti\'{c}$^{1,}$}
\email{vuletic@mit.edu}
\affiliation{
$^{1}$Department of Physics and Research Laboratory of Electronics, Massachusetts Institute of Technology, Cambridge, Massachusetts 02139, USA \\
$^{2}$Cavendish Laboratory, JJ Thompson Ave, Cambridge CB3 0HE, UK \\
$^{3}$Department of Chemistry and Chemical Biology, Harvard University, Cambridge, Massachusetts 02138, USA \\
$^{4}$Institute for Quantum Computing and Department of Physics and Astronomy, University of Waterloo, Waterloo, Ontario N2L 3G1, Canada 
}

\maketitle

In this supplement, we detail an iterative procedure to calculate slip probabilities $p_i$ from the experimentally measured fluorescence traces of the ion, with peak heights $f_i$.  We begin with a basic model, and then add measured corrections for imperfect experimental conditions.  We include a brief discussion of the measurement of the fluorescence peaks themselves.  Finally, we conclude with a comment on the calculation of the frictional energy dissipation shown in Fig.~4b of the main text.  Other details of the apparatus are discussed in Refs. \cite{karpa,science,velocity,aubry}.

\section{Extracting localization probabilities - basic model}

We consider a multistable potential landscape with local minima labeled $\alpha, \beta, \gamma,...$ from left-to right, with an ion localized in the left-most minimum $\alpha$.  As described in the main text, the landscape is the sum of a sinusoidal periodic potential and a harmonic trap potential.  Translating the harmonic trap to the right tilts the overall potential, causing minimum $\alpha$ to eventually become an inflection point, and the second-left-most minimum $\beta$ to become the new left-most minimum.  The appearance of the inflection point marks a slipping point: the ion is forced to slip, cool, and re-localize in the various remaining minima with some probability.  We make no assumption about the nature of this probability distribution.   

We can determine the probability distribution directly from ion fluorescence traces.  An ion's fluoresence peaks at a slipping point.  The height of the peak, averaged over many realizations of the experiment, is determined by the fraction of slip events occuring at that time; the more often an ion slips, the higher the corresponding peak.  The first peak, with height $f_1$, is always the largest because the ion is always initially prepared in the same global minimum.  We normalize all populations to $f_1$, which represents the full ion ensemble population; consequently, $f_1$ acts as a conversion between fluorescence and population.  After this first slip, the ion population localizes among the various empty minima: the new left-most minimum $\beta$ has population $p_A$, where $p_A$ denotes the probability of localizing into the left-most minimum. The next-left-most minimum $\gamma$ has population $p_B$, where $p_B$ is the probability of localizing into the next-left-most minimum, etc.

We continue to translate the harmonic trap as before.  After a translation of one lattice spacing $a$, the left-most minimum $\beta$, with ion population $p_A$, becomes an inflection point.  Its ion population slips, producing peak 2 with height $f_2 = f_1 \cdot p_A$.  The population of the remaining minima increase: the new left-most minimum $\gamma$, for example, now has population $f_2/f_1 \cdot p_A + p_B$, which will eventually slip and form the third fluorescence peak $f_3 = f_2 \cdot p_A + f_1 \cdot p_B$ \cite{footnote}.

Iterating this procedure over several slip events gives us a general relation for the peak heights:
\begin{equation}
f_i = f_{i-1} \cdot p_A + f_{i-2} \cdot p_B + f_{i-3} \cdot p_C + \cdots,
\end{equation}
where $f_{i-j} = 0$ if $i-j \leq 0$.  Thus, we can solve a set of coupled algebraic expressions to obtain the slip probability distribution $\{p_A,p_B,p_C\}$ from the measured height distribution $\{f_i\}$:
\begin{align}
f_2 &= f_1 \cdot p_A, \\
f_3 &= f_2 \cdot p_A + f_1 \cdot p_B, \\
f_4 &= f_3 \cdot p_A + f_2 \cdot p_B + f_1 \cdot p_C.
\end{align}
Note that we are perfectly constrained; there is no need to add the constraint $p_A+p_B+p_C=1$.  This condition, however, can be used to verify errors in our extraction of proabilities arising, for example, from thermal hopping.  Solving this set of Eqs.~2-4 results in Eq.~2 of the main text.

\section{Corrections to the basic model}

Eq.~1 is a good working model for extracting slip probabilities, but it is only valid for perfect experimental conditions.  The following are two corrections that we make to our basic model to account for experimental imperfections.

\subsection{Misinitialization correction}

The basic model assumes that $100\%$ of the ion ensemble population is initially localized in minimum $\alpha$ and slips together to form the first fluorescence peak $f_1$.  In the experiment, we initially localize the ion in the global minimum (at high $\eta$, this is the central of multiple minima), and a small fraction of the total ion population often spills over equally into the two minima adjacent to the global minimum.  The small misinitialization in the left-adjacent minimum, in particular, can be observed as a tiny fluorescence bump that precedes the first large fluorescence peak.  By measuring the bump fluorescence $f_{bump}$, we can determine the fraction of the total ensemble that was misinitialized and account for its contribution to the adjacent minimum.

The misinitialization correction changes the slip probability distribution by less than $10 \%$.

\begin{figure}[htbp]
	\begin{center}
		\includegraphics[width=0.45\textwidth]{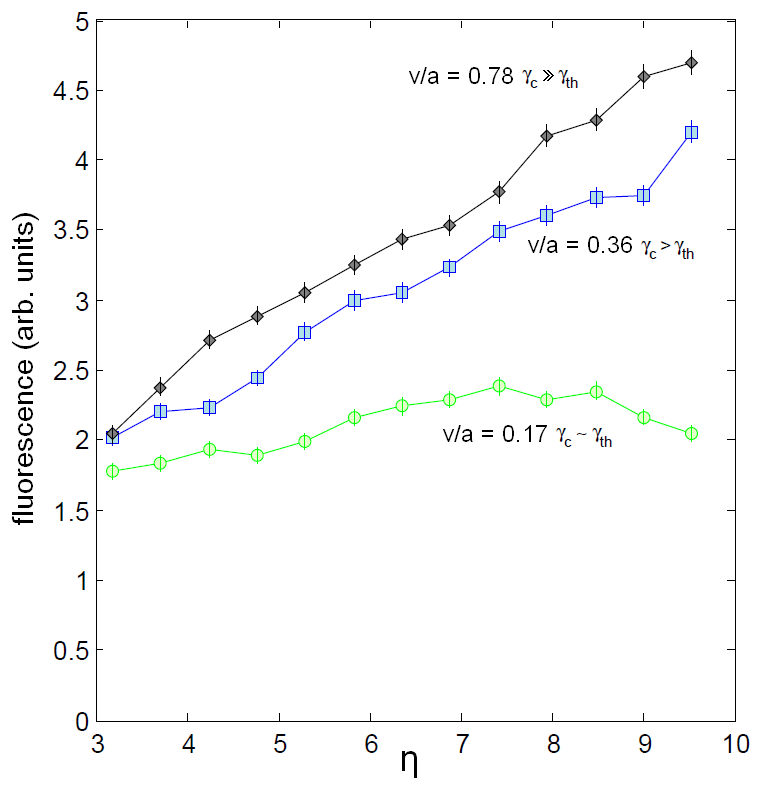} 
		\caption{First peak fluorescence $f_1$ across a range of values of the corrugation parameter $\eta$ and trap translation speed $v$.  The fastest curve (black diamonds), which is least subject to thermal hopping, is used as a reference.  Note the increase of fluorescence with $\eta$, due to the dependence of $\eta$ on optical-lattice intensity.  Any deviation from a monotonic increase with $\eta$ is due to thermal hopping.}
	\end{center}
\end{figure}

\subsection{Thermal hopping correction}

The basic model also neglects thermal hopping.  An ion, due to its finite temperature, has a non-negligible probability of being thermally excited and localizing in another minimum, without producing an (averaged) fluorescence peak at a well-defined slip time.  Such an ion will not contribute to the fluorescence peak that forms at the true slipping point and thus will lead to the assignment of an incorrect localization probability.  To determine the fraction of ions that prematurely hops, we compare the first peak fluorescence $f_1$ for a given $\eta$ across different velocity datasets (Fig.~1).  The dataset with the fastest translation ($v/a = 0.78$ $\gamma_c \gg \gamma_{th}$) can be used as a reference, as it is in a regime in which thermal hopping is negligible \cite{velocity}.  With this dataset as a reference, we determine the fraction of the ion population that hops without contributing to the fluorescence peak, and we can account for this population's contribution to the adjacent minimum.

The thermal hopping correction changes the slip probability distribution by less than $20 \%$ for the slowest translation speed $v/a = 0.17 \gamma_c$ and less than $10 \%$ for the middle translation speed $v/a = 0.36 \gamma_c$.  We note that while we can correct for thermal hopping out of the left-most minimum just prior to a slip (as discussed above), we do not have a way of rigorously correcting for thermal hopping amongst the other minima, in between slips.  It is this latter effect that is responsible for deviation from the model in Fig.~3a of the main text.

\begin{figure}[htbp]
	\begin{center}
		\includegraphics[width=0.45\textwidth]{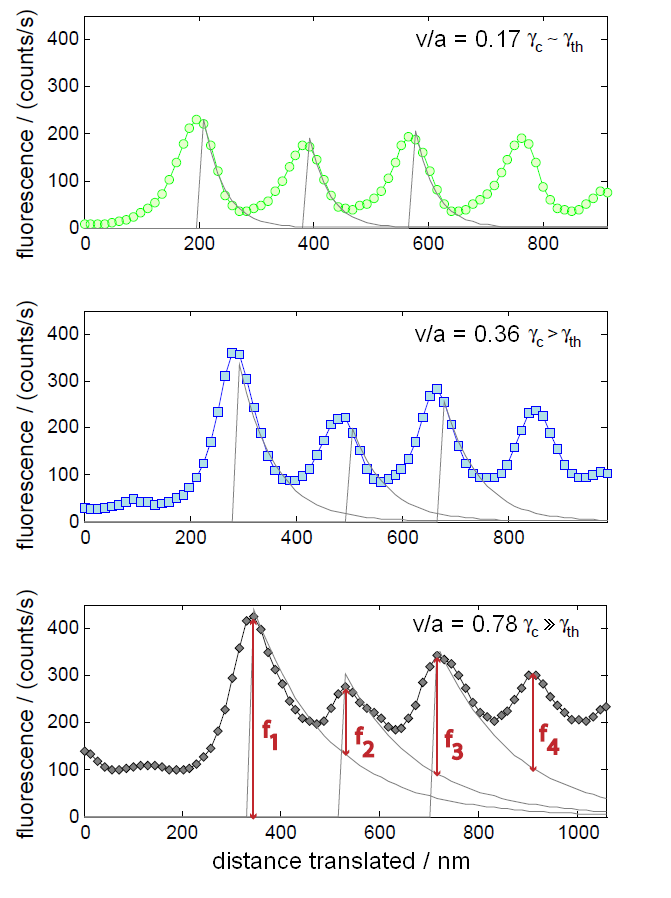} 
		\caption{Fluorescence traces, and exponential fits, for $\eta = 7.95$ across three different velocities, as a function of the harmonic trap translation.  At the slowest velocity, thermal hopping is pronounced, leading to smoothing of any difference in peak heights.  At the highest velocity, recooling causes a pronounced exponential fluorescence decay that must be subtracted.  For the middle velocity, neither thermal hopping nor recooling effects are especially pronounced.}
	\end{center}
\end{figure}

\section{Correction of fluorescence peak heights due to finite cooling rate}

\begin{figure}[htbp]
	\begin{center}
		\includegraphics[width=0.45\textwidth]{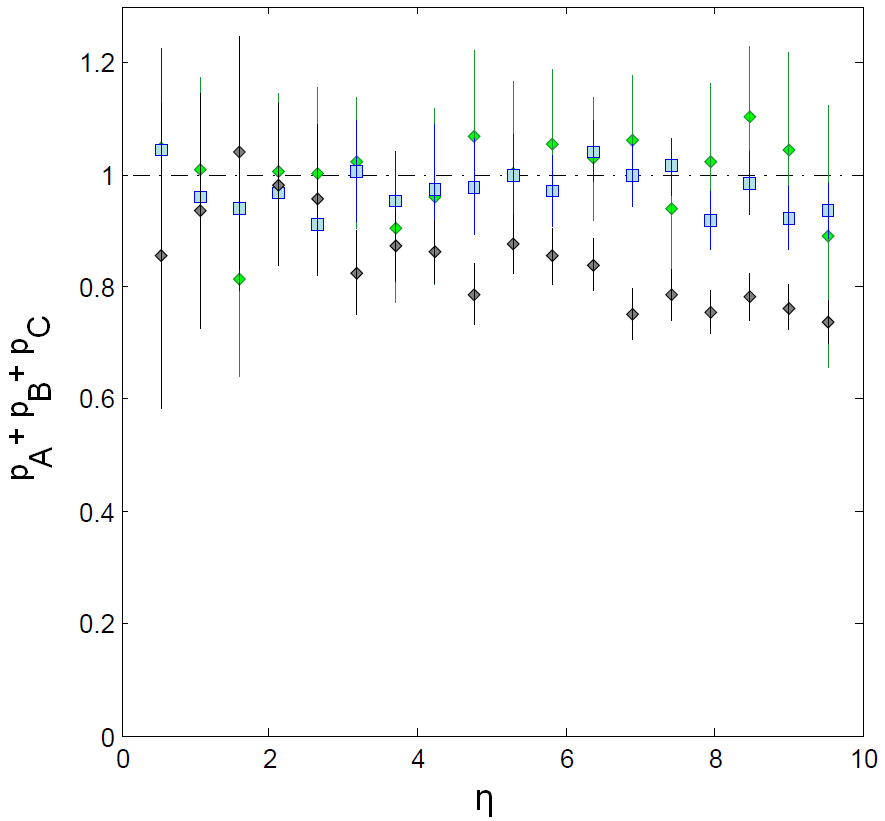} 
		\caption{Summed probabilities as a function of the corrugation parameter $\eta$ across three different transport velocities ($v/a = 0.17$ $\gamma_c$ for green circles, $v/a = 0.36$ $\gamma_c$ for blue squares, $v/a = 0.78$ $\gamma_c$ for black diamonds).  The fastest dataset requires normalization.}
	\end{center}
\end{figure}

Finally, we note that measuring the heights of the fluorescence peaks is complicated by the finite recooling rate $\gamma_c$ of the system.  Fluorescence from a slip may still be exponentially decaying at the time of the following slip.  If one does not take this effect into account, the following slip's peak height will be overestimated.  This effect is especially pronounced at high velocity, where $v/a$ approaches $\gamma_c$.  Thus, we fit each peak's fluorescence to a decaying exponential and subtract the value of this exponential when measuring peak heights (see Fig.~2).  We note that this method is imperfect, as seen in Fig.~3, where the highest velocity dataset shows that $p_A + p_B + p_C < 1$.  To correct for this, we re-normalize all datasets to ensure $p_A + p_B + p_C = 1$.  This correction changes the probability distribution by less than 5$\%$ for the two slower velocity datasets, but amounts to a less than 20$\%$ correction for the fastest dataset $v/a = 0.78 \gamma_c$.

We also see effects from the finite thermal hopping rate $\gamma_{th}$, which tends to smooth out any difference in peak heights.  This effect is most pronounced at the lowest velocity and is partially corrected, as discussed above. 

\section{Frictional Energy Dissipation}
The frictional energy $U_{diss}$ dissipated by our laser cooling system after each slip event can be determined from detailed knowledge of the energy landscape:
\begin{align*}
U_{diss} &= p_A \Big( V_0 - V_A(\tau) \Big)\\ \numberthis \label{eqn}
	     &+ \frac{p_B}{2} \Big( V_0 - V_B(\tau) \Big)\\
	     &+ \frac{p_C}{3} \Big( V_0 - V_C(\tau) \Big). 
\end{align*}
Here, $V_0$ is the energy of the ion at the slipping time, and $V_i(\tau)$ is the energy of the $i$th minimum at the localization time $\tau$.  These energy values are calculated from the Prandtl-Tomlinson model (Eq.~1 of the main text).